\documentstyle[seceq,preprint,epsbox]{jpsj}

\title
{\bf Phase Transitions in Bilayer 
  Heisenberg Model with General Couplings} 
\author
{ 
Yasuhiro {\sc Matsushita}\footnote{E-mail: 
matusita@grad.ap.kagu.sut.ac.jp}, 
Martin P. {\sc Gelfand}$^{1}$ 
and Chikara {\sc Ishii}}

\inst
{Department of Physics, Faculty of Science,
Science University of Tokyo \\
$^1$Department of Physics, 
Colorado State University, 
Fort Collins, Colorado 80523}

\recdate
{
}

\abst
{ The ground state properties and phase diagram of 
the bilayer square-lattice Heisenberg model 
are studied in a broad parameter space of 
intralayer exchange couplings, 
assuming an antiferromagnetic coupling between 
constituent layers. 
In the classical limit, the model exhibits 
three phases: 
two of these are ordered phases 
specified by the ordering wave vectors 
$(\pi,\pi;\pi)$ and $(0,0;\pi)$, 
where the third component of each indicates antiferromagnetic 
orientation between layers, while another is a canted phase, 
stabilized by competing interactions. 
The effects of quantum fluctuations in the model 
with $S=1/2$ have been explored 
by means of dimer mean-field theory, 
exact diagonalization 
of $2\sqrt2\times2\sqrt2\times2$ clusters, 
and high-order perturbation expansions 
about the interlayer dimer limit. }
\kword
{
bilayer Heisenberg model, dimer mean-field theory, 
exact diagonalization, series expansion, phase diagram} 

\begin{document}
\sloppy
\maketitle

\section{Introduction}
 Quantum magnets in two dimensions have 
been a fascinating topic of numerous studies 
over the past decade. 
The interest has primarily focused on 
the competition between long-range magnetic order 
and novel disordered phases at $T=0$ as a result of 
enhanced quantum fluctuations. 
In this paper, we consider a bilayer square-lattice Heisenberg model 
with general 
nearest-neighbor exchange couplings 
in each layer and antiferromagnetic 
coupling between corresponding 
sites of each layer, described by the Hamiltonian 
\begin{equation}
 H=J_{1}\sum_{\langle i,j \rangle}{\bf S}_{1,i}{\cdot}{\bf S}_{1,j}
+J_{2}\sum_{\langle i,j \rangle}{\bf S}_{2,i}{\cdot}{\bf S}_{2,j}
+\sum_{i}{\bf S}_{1,i}{\cdot}{\bf S}_{2,i}\ ,
\end{equation}
where the interlayer coupling is taken as the unit of energy.

A particular limit of the model in which $J_1=J_2\equiv J>0$
has attracted special attention for several reasons. 
Several cuprate superconductors contain 
$\rm CuO_{2}$ bilayers, so that model may be 
relevant to their magnetic properties\cite{Millis} --- and it is
certainly an appropriate model for the magnetic
properties of their parent compounds.
Its phase diagram has been studied 
numerically by quantum Monte Carlo \cite{Sandvik} and
perturbation expansions about $J=0$ \cite{Hida,Gelfand1,Weihong}, 
and analytically using linear spin-wave theory\cite{Matsuda}, 
Schwinger boson mean-field theory\cite{Ng}, 
and an alternative bosonic mean-field calculation\cite{Morr}. 
This one-parameter subspace of the Hamiltonian (1.1) has two 
well known limits.
For $J\gg1$, the model describes a pair of 
two-dimensional Heisenberg antiferromagnets, 
weakly interacting with each other. 
In the isolation limit, each would be in the ordered state 
at $T=0$ and would possess Goldstone modes related 
to spontaneous breakdown of spin-rotational symmetry.
Upon weak coupling, their sublattice magnetizations 
are phase-locked and half of the Goldstone modes acquire gaps. 
For $J\ll1$, the ground state is very close to the ``dimer'' state, 
the product of spin singlets 
composed of adjacent interlayer pairs of spins.  
This state has a gap of order unity. 
Thus the system has a singlet ground state 
both for $J\gg1$ and $J\ll1$, 
and one can anticipate that there should be
a critical value $J_c$ of the intralayer coupling 
at which a continuous transition between the 
two phases is expected; furthermore, 
we can anticipate that this transition belongs to 
the universality class of the classical 
$d=3$ Heisenberg model \cite{Chakravarty}. 
The various numerical studies all concur that $J_c\approx0.394$. 

We were led to consider the more general Hamiltonian (1.1)
on the following grounds. 
First of all, for $J_2=0$ the model is related to the ``Kondo necklace'' model 
introduced by Doniach\cite{Doniach} 
as a simple model to describe the competition 
between the intra-atomic Kondo coupling (favoring local 
singlet states) and the inter-atomic RKKY interaction 
(favoring magnetic order and local high-spin states) 
in heavy fermion systems. 
The model for $J_2=0$ may also be viewed as 
the symmetric Hubbard-Kondo lattice model 
in the limit of strong on-site repulsion $U$ 
between conduction electrons\cite{Igarashi, Shibata}. 
In the conventional symmetric Kondo lattice model ($U=0$) 
on the square lattice
with Kondo coupling $J_{K}$, 
it is known that a transition from spin-liquid to antiferromagnetic 
phase takes place when the transfer energy $t$ 
is increased beyond $t_c\approx 0.7 J_K$\cite{ZPS}. 
Our calculation in \S{3.2} (the case $\theta=0$) 
indicates that there is such a transition at 
$t_c\approx0.42\sqrt{UJ_K}$ (for $U\gg J_K$): 
The antiferromagnetic order is suppressed 
by the strong on-site repulsion. 
This result is consistent with a finding by Shibata 
{\it et al.}~\cite{Shibata} for the one-dimensional 
symmetric Hubbard-Kondo lattice model, that
the spin gap and inverse spin-spin correlation length 
are increasing functions of $U$. 
For $J_{2} \neq 0$, our results describe 
the effects of direct interactions between 
localized spins on the stability of the Kondo-singlet 
phase, which is neglected in the most analyses\cite{directH}. 
As would be naively expected, 
the antiferromagnetic direct interactions 
are shown to favor the long-range 
ordered phase over Kondo singlets, while the opposite 
is true for (weak) ferromagnetic direct interactions. 

When $J_1$ and $J_2$ have opposite signs, 
the model is uniformly frustrated:
The product of couplings around any four-spin plaquette that
encompasses both layers is negative. 
Uniform frustration has been shown to lead 
to exotic magnetic behavior in many cases,
both for theoretical models \cite{majumdarghosh} and real materials 
\cite{Ramirez}. 
Therefore it would be of great interest to see what are expected 
in the present relatively simple model. 
We find no evidence for any exotic magnetic singlet phases,
such as a spontaneously dimerized or plaquettized phase.

We have carried out several types of analyses 
in order to identify the phases and phase transitions 
exhibited by the Hamiltonian (1.1). 
The classical ground state phase diagram is 
first considered (in \S{2}). 
Then quantum effects for $S=1/2$ spins are included, 
and it is clearly shown that a spin-disordered phase 
(the aforementioned dimer phase) appears 
in the vicinity of $J_1=J_2=0$. 
We examine the instabilities of this 
phase by means of mean-field theory, small-system
exact diagonalization (in \S{3.1}), 
and higher-order perturbation 
expansions for susceptibilities 
and excitation spectra (in \S{3.2}). 
The last technique also allows us to determine the 
spin-wave velocity in the magnetically ordered states 
in the vicinity of the phase boundary. 
The last section (\S{4}) is devoted to summary and discussion. 
In particular, Sachdev and Senthil \cite{Sachdev} recently considered
a quantum rotor model which is closely related to the model (1.1),
and in \S{4} we will compare their results with ours.

\section{The classical limit}

For our purpose in the following, 
it is instructive to consider 
the ground state phase diagram 
in the classical limit, 
where the operators $\bf S$ in the Hamiltonian (1.1) 
are replaced by classical vectors.
By noting that the interchange
$J_1 \leftrightarrow J_2$ simply 
corresponds to relabeling the layers, 
we only need to consider the half space 
$J_1\ge J_2$ of the $J_1$-$J_2$ plane. 

One can easily determine
the ground states in the part of parameter space 
where the two intralayer couplings have the same sign.
When $J_1$ and $J_2$ are both positive the ground state
is composed of a pair of N\'{e}el ordered layers 
with an ordering wave vector $(\pi,\pi;\pi)$, 
where the third component indicates the antiferromagnetic 
orientation between layers. 
When $J_1$ and $J_2$ are both negative
the ground state is composed of 
two ferromagnetically (but oppositely) ordered layers 
with an ordering wave vector $(0,0;\pi)$. 
In each case, both intra- and interlayer energies 
of the system can be minimized simultaneously. 

The $(\pi,\pi;\pi)$ phase and $(0,0;\pi)$ phase 
both extend into the quadrant $J_{1}>0>J_{2}$, 
bordered on a phase which will be referred to 
as the weakly ferromagnetic phase (WF). 
This phase is composed of a canted N\'{e}el-ordered and
a canted ferromagnetic layers, as is shown in Fig.~1. 
The direction of the staggered moment in the antiferromagnetically
coupled layer is perpendicular to the uniform moment 
on the ferromagnetic layer.
The canting angles $\psi$ and $\phi$ in the layers satisfy 
\begin{equation}
\sin\psi=\xi_2\sqrt{ (1-{\xi_1}^2)/(1-(\xi_1\xi_2)^2)}
\end{equation}
and 
\begin{equation}
\sin \phi =\xi_1\sqrt{ (1-{\xi_2}^2)/(1-(\xi_1\xi_2)^2)}
\end{equation}
with $\xi_i$ defined by
\begin{equation}
\xi_i=4J_i\left(-1+\sqrt{1+1/\left(16J_{1}J_{2}\right)}\right) \ .
\end{equation}
When $J_1$ and $J_2$ are large, the canting angles behave like 
$\psi\approx 1/(8J_1)$ and $\phi\approx 1/(8J_2)$. 

The boundary between the $(0,0;\pi)$ phase
and the WF phase is the line $1/J_{1}+1/J_{2}=8$,
while the boundary between the $(\pi,\pi;\pi)$ phase
and the WF phase is the line $1/J_{1}+1/J_{2}=-8$.
In both cases the phase transition is continuous.
The classical phase diagram is plotted
in Fig.~2.  

\begin{center}
\psbox[scale=0.4]{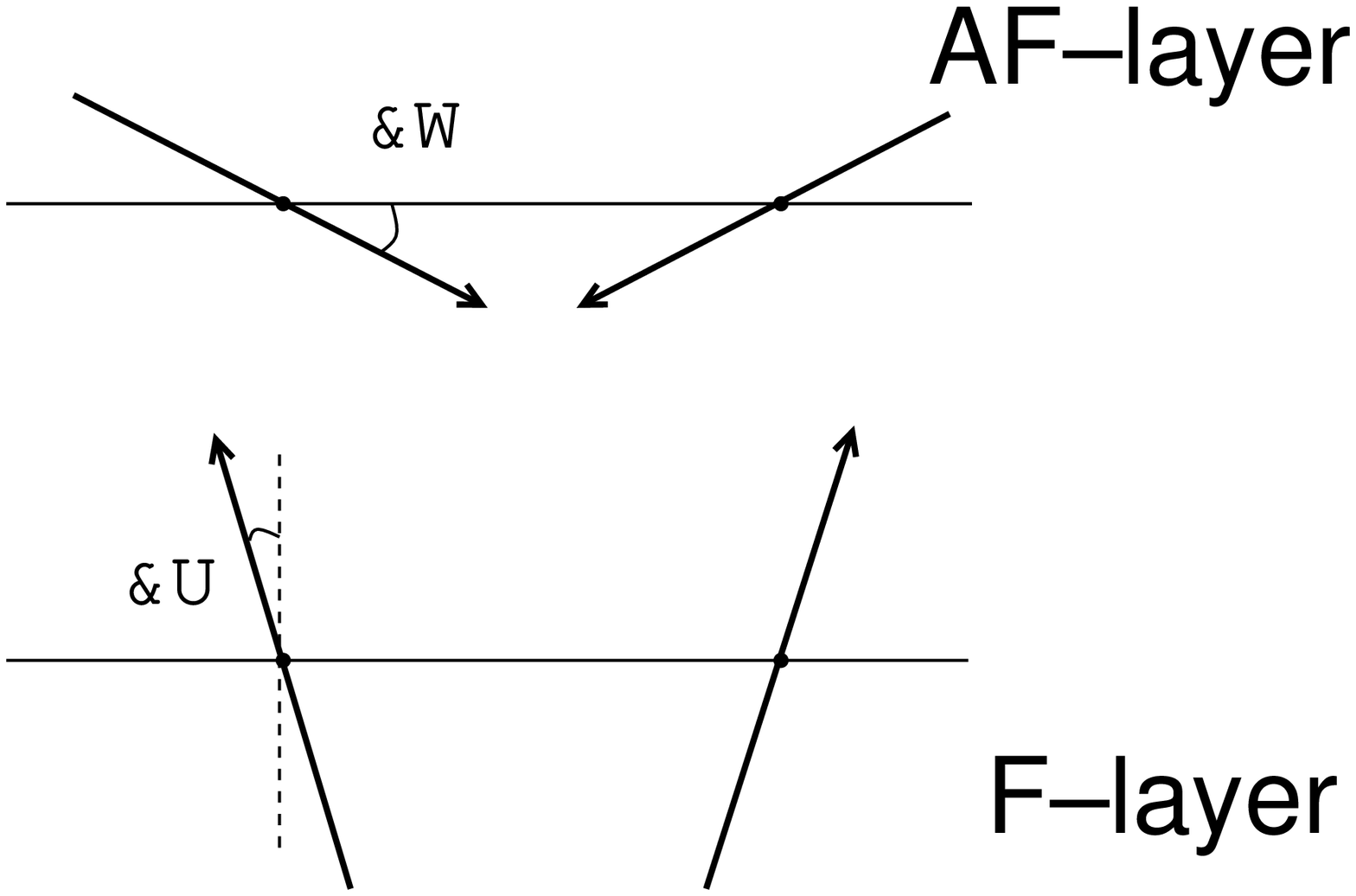}
\begin{quote}
\begin{footnotesize} 
Fig.~1. \hspace{6pt} Canting angles 
$\psi$, $\phi$ in the antiferromagnetic (AF) 
and ferromagnetic (F) layers respectively. 
\end{footnotesize}
\end{quote}
\end{center}

\begin{center}
\psbox[scale=0.4]{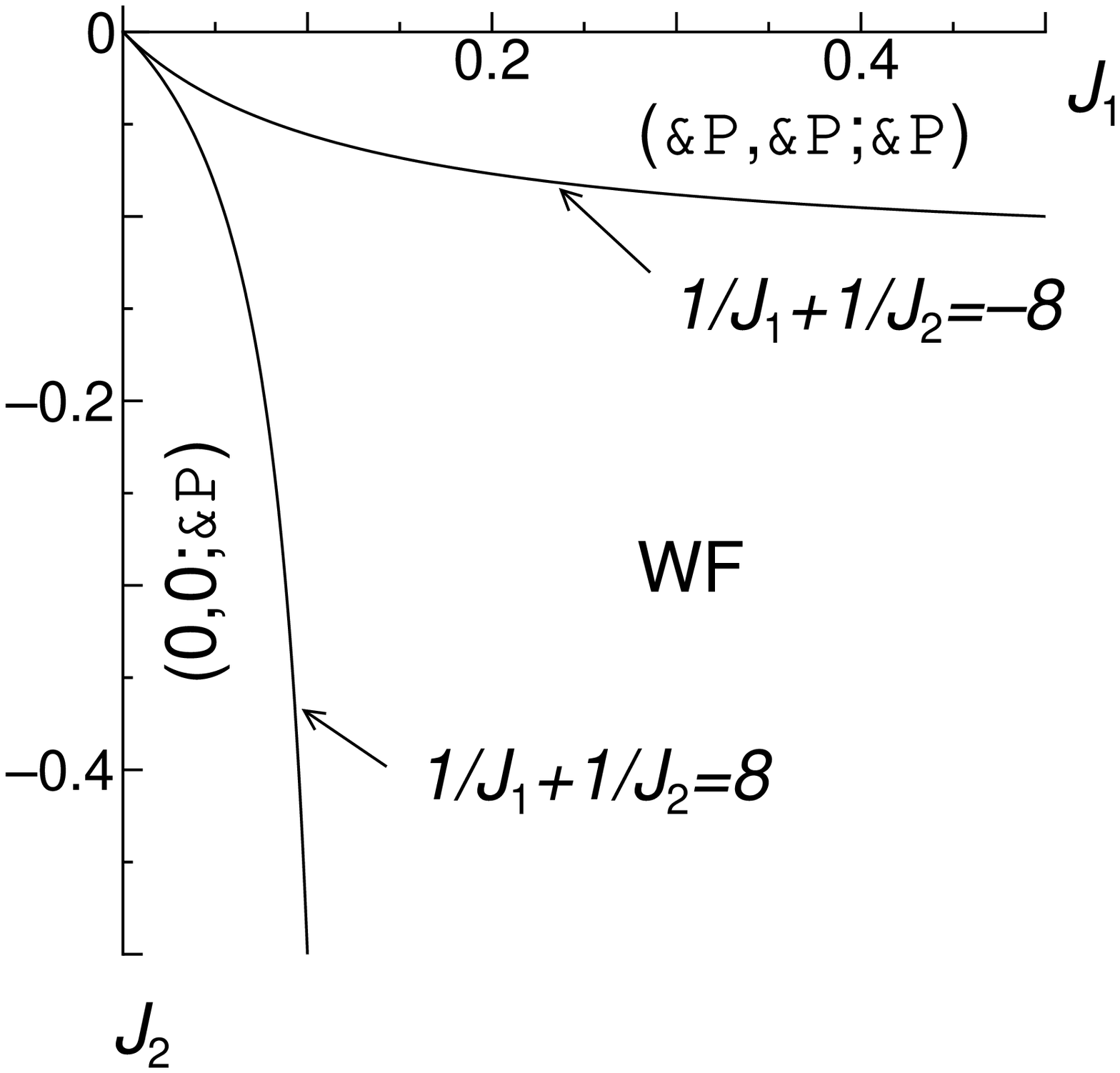}
\begin{quote}
\begin{footnotesize} 
\begin{center}
Fig.~2. \hspace{6pt} Phase diagram 
in the classical spin limit
($S\rightarrow\infty$). 
\end{center}
\end{footnotesize}
\end{quote}
\end{center}
\vspace{8pt}

\section{Phase diagram for the quantum model} 

\subsection{Simple considerations}

There is at least one region in parameter space where 
the classical phase diagram, Fig.~2, should be modified 
for the quantum spin model with $S=1/2$. 
When $J_1$ and $J_2$ are both small, 
the ground state is approximately a product of 
interlayer-dimer singlets and does not 
exhibit long-range order in its spin-spin correlation function.
Since there is an energy gap of 1 (the interlayer coupling) 
for $J_1=J_2=0$, we can expect that the ``dimer'' phase is 
stable in a region enclosing the origin with diameter of order unity.  
Let us try to make this statement 
more precise, by means of relatively simple calculations.

An estimate of the boundaries between the dimer phase and 
long-range ordered phases [$(\pi,\pi;\pi)$, $(0,0;\pi)$, WF]
can be obtained from dimer mean-field theory.
One considers the following two-site Hamiltonian 
\begin{equation}
  {\bf S}_1 \cdot {\bf S}_2 - h_1S_1^z - h_2S_2^z -g_1S_1^x -g_2S_2^x
\end{equation}
and imposes the self-consistency conditions 
\begin{equation}
h_1=4J_1\langle S_1^z\rangle\qquad h_2=4J_2\langle S_2^z \rangle
\qquad g_1=g_2=0
\end{equation}
for the $(\pi,\pi;\pi)$ phase, or
\begin{equation}
h_1=-4J_1\langle S_1^z\rangle\qquad h_2=-4J_2\langle S_2^z \rangle
\qquad g_1=g_2=0
\end{equation} 
for the $(0,0;\pi)$ phase, or
\begin{equation}
h_1=4J_1\langle S_1^z\rangle\qquad h_2=4J_2\langle S_2^z \rangle \qquad
g_1=-4J_1\langle S_1^x\rangle\qquad g_2=-4J_2\langle S_2^x \rangle
\end{equation}
for the WF phase.  In the above self-consistency conditions,
the angular brackets denote ground-state averages. 
 
As usual in mean-field theory, 
one finds that $h_1=h_2=g_1=g_2=0$ is the only solution, 
and hence the dimer phase is stable, 
for a small interdimer couplings. 
It is straightforwardly shown that the dimer phase
is unstable against the $(\pi,\pi;\pi)$ phase for
$J_1 +J_2>1/2$, and the $(0,0;\pi)$ phase for
$J_1 +J_2<-1/2$; in both cases the transitions are continuous. 
Much more effort is needed to determine the boundary of the WF phase, 
and we have not been able to derive any analytical results. 
Instead, we have carried out numerical calculation, 
solving the mean-field equations iteratively
(starting with a set of prescribed magnetizations to 
calculate the fields and solving the two-site
Hamiltonian for the magnetizations, repeating 
these process until convergent results are obtained). 
The resulting phase boundaries are shown in Fig.~3. 
An unphysical feature of our results is that the ground-state energy 
exhibits a discontinuous decrease on entering the WF phase. 
This might be due to the WF solutions which 
are not stable under iteration, 
so that we have underestimated the 
domain of stability of the WF phase. 

We have also considered another type of mean-field calculation, 
which was applied to the special case $J_1=J_2$
by Chubukov and Morr\cite{Morr}.  Their calculation can
be readily generalized. 
Let us omit all details, which are clearly described in their paper,
and just describe the results.  By rewriting the Hamiltonian
in terms of three types of bosonic operators, 
corresponding to the three triplet excited states 
of an $S=1/2$ dimer, and keeping only the 
quadratic terms, one ends up with a three-fold degenerate
excitation spectrum of the form
\begin{equation}
\epsilon({\bf q})=\sqrt{1+(J_1+J_2)(\cos q_x + \cos q_y)}\ .
\end{equation}
(The in-plane lattice spacing is taken as the unit of length.)
At small intralayer couplings the spectrum has a gap, characteristic 
of the dimer phase.  When $|J_1+J_2|$ is increased beyond $1/2$,
some of the excitation
energies pass through zero onto the imaginary axis, signaling
an instability of the dimer phase.  This analysis agrees completely
with the dimer mean-field calculations for the transitions to
the $(0,0;\pi)$ and $(\pi,\pi;\pi)$ phases, as described above.  
It also yields the result that the spin-wave velocity on
both of those phase boundaries is given simply by
$c=1/2$.
However, this analysis does not appear capable of describing the
transition to the WF phase. 

\begin{center}
\psbox[scale=0.4]{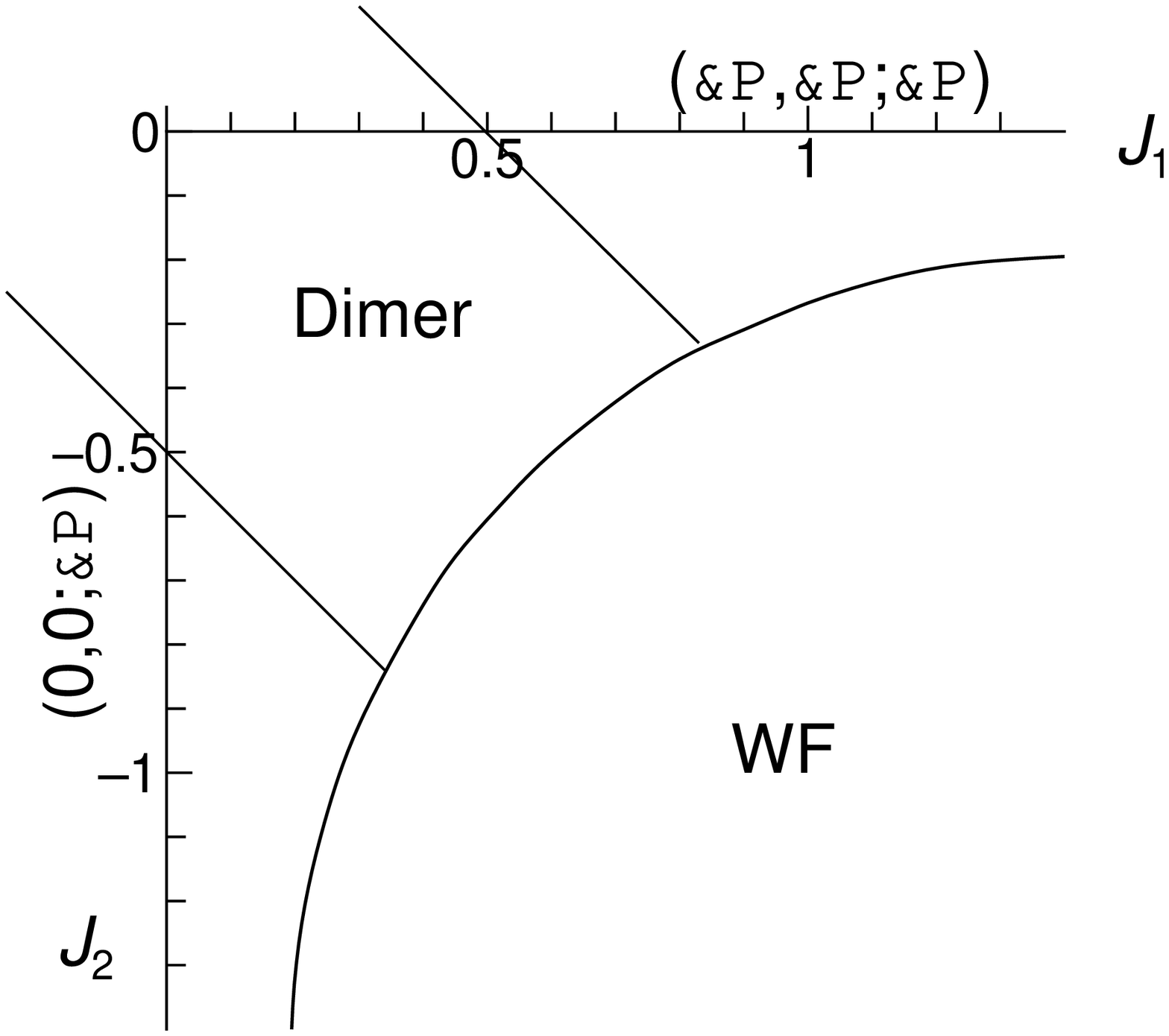}
\begin{quote}
\begin{footnotesize} 
\begin{center}
Fig.~3. \hspace{6pt} Phase diagram 
for $S=1/2$ spins from dimer mean-field theory. 
\end{center}
\end{footnotesize}
\end{quote}
\end{center}
\vspace{10pt}

Finally, we have 
carried out finite-size exact diagonalization 
studies on $2\times2\times2$ 
and $2\sqrt2\times 2\sqrt2 \times 2$ systems with periodic boundary
conditions\cite{Nishimori}. 
In Fig.~4 the low-lying energy levels are plotted as functions
of $J\equiv|J_{1}|=|J_{2}|$ for the three distinct choices
of the signs of couplings. 
One finds that when the intralayer couplings 
have the same sign the ground state 
is always a singlet and when the couplings have opposite
signs there are level crossings to high-spin states as
the magnitude of the couplings increases.  The latter 
is associated with the transition from the dimer phase 
to the WF phase; in the WF phase the ground state
has a net moment and therefore must have nonzero spin.
We will have more to say about the finite size calculations 
in the following subsection. 
\begin{center}
\psbox[scale=0.35]{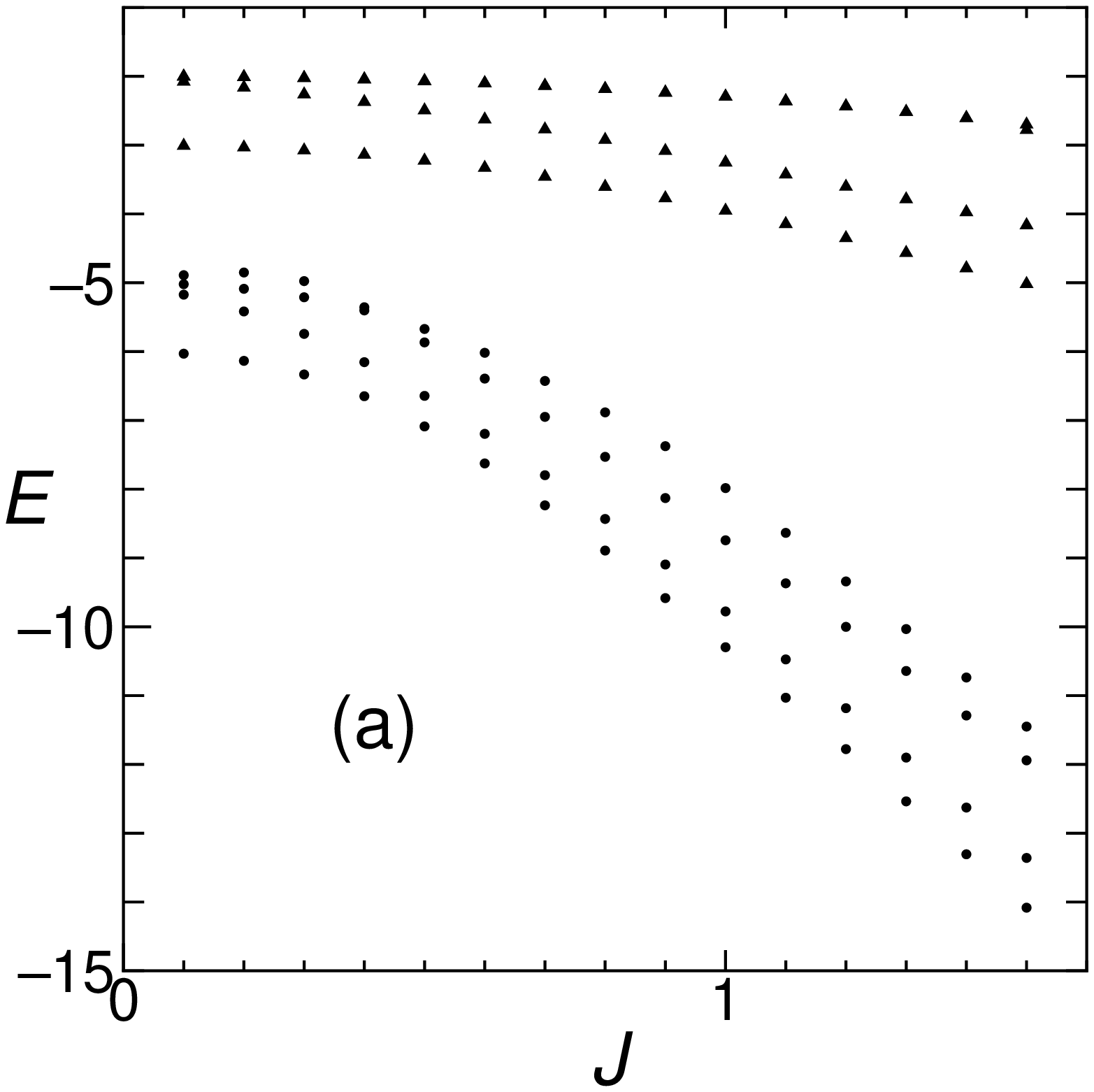}
\psbox[scale=0.35]{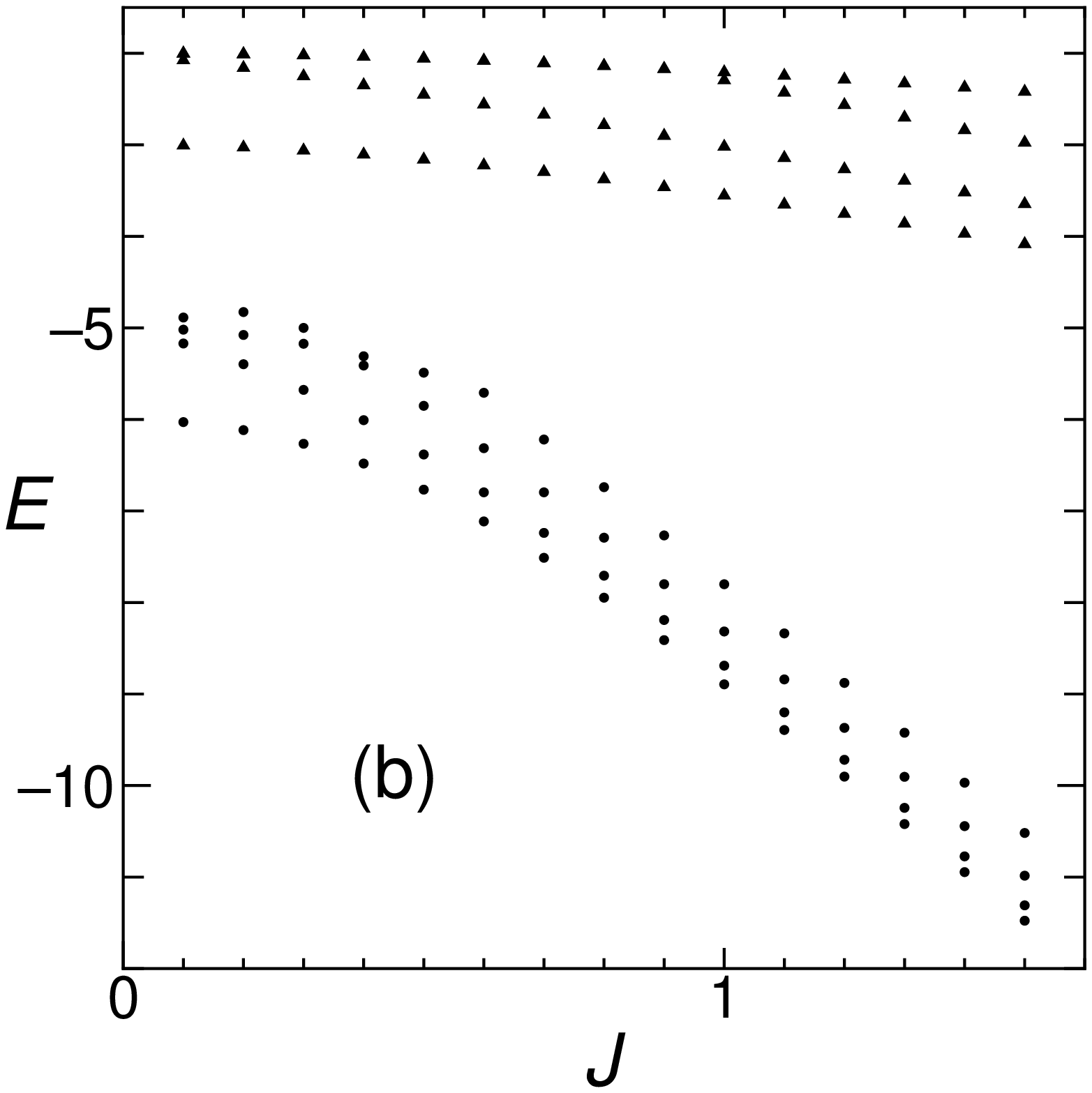}
\psbox[scale=0.35]{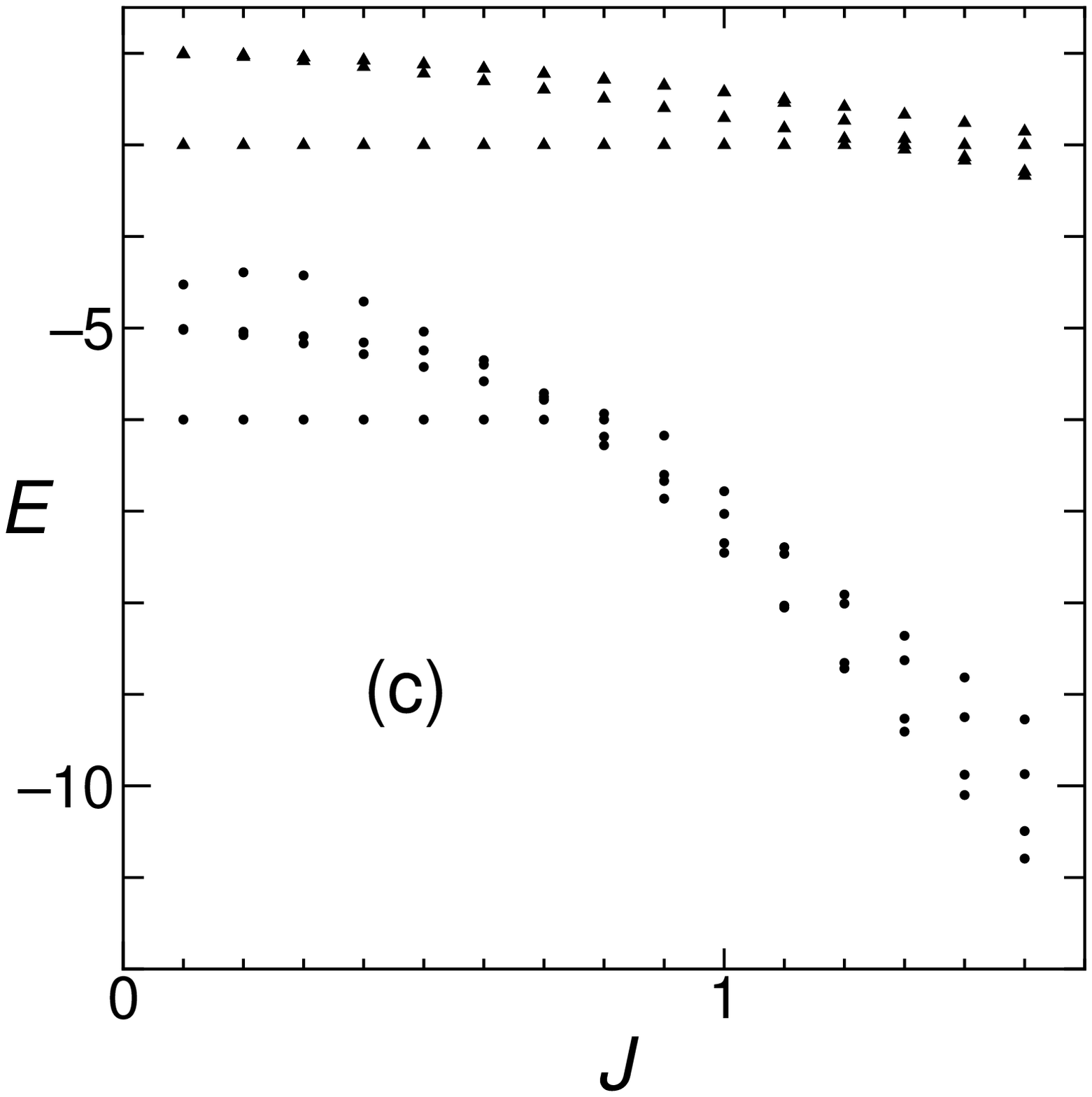}
\begin{quote}
\begin{footnotesize} 
Fig.~4. \hspace{6pt} Low-lying energy levels of 
$2{\times}2{\times}2$ (triangles) and 
$2\sqrt2{\times}2\sqrt2{\times}2$ (dots) clusters 
as functions of $J\equiv |J_{1}|=|J_{2}|$
: (a)$J_{1}>0, J_{2}>0 $, (b)$J_{1}<0, J_{2}<0 $, 
(c)$J_{1}>0, J_{2}<0 $. 
\end{footnotesize}
\end{quote}
\end{center}
\vspace{10pt}

\subsection{Series expansions and extrapolations}

In order to obtain more quantitative estimates
of the domain of stability of dimer phase 
and the spin-wave velocity in the magnetically ordered phases 
adjacent to those phase boundaries, we have carried out 
perturbation expansions in powers of the intralayer
couplings about the pure interlayer dimer Hamiltonian.

For convenience, the intralayer 
couplings are parametrized by 
\begin{equation}
 J_{1}=\lambda\cos\theta \qquad J_{2}=\lambda\sin\theta \ . 
\end{equation} 
Expansions in $\lambda$ can then be carried out at fixed $\theta$, 
rather than carrying out a double expansion in powers
of $J_1$ and $J_2$ simultaneously.
Note that only expansion for $\theta$ 
in the range $-\pi/4$ to $\pi/4$ is needed, 
since {\it negative\/} values of $\lambda$ 
correspond to $\theta$ in the range $3\pi/4$
to $5\pi/4$, which then maps to $-\pi/4$ to $-3\pi/4$
under the interchange $J_1\leftrightarrow J_2$. 
In this way full $2\pi$ coverage is obtained by 
carrying out calculations only for a quarter of the circle,
which were done at intervals of $\pi/16$.
Series expansions 
for several susceptibilities $\chi({\bf q})$ at $T=0$ 
and the Fourier transform of the triplet elementary 
excitation spectrum $\epsilon({\bf q})$ 
were performed up to the order ${\lambda}^{8}$ 
using connected cluster methods \cite{Gelfand2,Gelfand1}. 

The values of $\lambda$ corresponding to 
continuous transitions out of the dimer phase
can be accurately estimated by 
constructing differential approximants\cite{Fisher} to the
susceptibility and the gap series associated with
the ordering wave vector. 
The leading critical behavior of 
these quantities is described by 
$\chi\sim (\lambda_{c}-\lambda )^{-\gamma}$ 
and $\epsilon\sim (\lambda_{c}-\lambda )^{\nu}$.
The dimer-$(\pi,\pi;\pi)$ and dimer-$(0,0;\pi)$
transitions should lie 
in the universality class of 
the $d=3$ classical Heisenberg model, 
and we can expect $\gamma\approx 1.4$ and $\nu\approx 0.71$\cite{Ferer} 
along those critical lines. 
If the dimer-WF transition is continuous, the naive hypothesis
is that it would lie in the $d=3$ classical $SO(3)$ class, 
as has shown to be the case for canted 
magnets by Kawamura \cite{kawamura}, 
for which $\gamma\approx 1.1$ and $\nu\approx0.53$. 
On the other hand, if the dimer-WF transition is of the first-order,
one would expect the linearly vanishing gap, since the transition
would be a result of a level crossing to a state of different symmetry.
The value of $\lambda$ at which the gap vanishes would be regarded as 
an upper bound on the actual phase boundary, 
since the excitations might have attractive interactions 
(and as we will see below, 
this is the case in the relevant parameter regime). 

In order to organize the following discussion, 
it is useful to divide the parameter space into three regions: 
region I, with $0\leq \theta \leq \pi/4$; 
region II, with $-3\pi/4 \leq \theta \leq -\pi/2 $, and
region III, with $ -\pi/2 < \theta < 0 $ . 

In region I, the dimer phase is adjacent to 
the $(\pi,\pi;\pi)$ phase, and 
a continuous transition between these phases is 
expected as in the conventional bilayer Heisenberg 
antiferromagnet ($\theta=\pi/4$). 
Shown in Fig.~5 is the phase diagram based on biased 
differential approximants to the expansion series 
$\chi(\pi,\pi;\pi)$ and $\epsilon(\pi,\pi)$. 

Note that without biasing, the critical exponents
derived from approximants show good agreement with the 
anticipated values, but the biasing allows for more
precise estimates of critical couplings than would be
possible otherwise.  The uncertainties in the critical 
value for $\lambda$ estimates are approximately 0.002. 

In region II, there is a continuous 
transition from the dimer phase to the $(0,0;\pi)$ 
phase, and again the critical exponents $\gamma$ and $\nu$ 
agree well with the anticipated values. 
Critical points derived from biased differential
approximants are shown in Fig.~5. 

 It is to be noted that the inequality $|J_1+J_2|>1/2$ 
along these two lines of continuous transitions is held, 
indicating that the mean-field theory underestimates 
the domain of stability of the dimer phase. 
This is entirely plausible, since mean-field theory neglects 
the effects of quantum fluctuations, which stabilize the dimer phase. 
Furthermore we note that these two lines are not 
mirror images of one another. 
The dimer-$(\pi,\pi;\pi)$ line 
is slightly but noticeably curved. The curvature
suggests that quantum fluctuations suppress N\'eel order
more strongly when the coupling between dimers is
evenly distributed between the layers than when the coupling
is concentrated in one layer. 
On the other hand, the critical dimer-$(0,0;\pi)$ line 
is almost as straight as is the case in dimer mean-field theory. 
We have not an clear explanation for it, 
but it is worth noting another point which shows 
that the $(\pi,\pi;\pi)$ and $(0,0;\pi)$ phases are 
affected differently by the quantum
fluctuations:  Within linear spin-wave theory, the spin-wave
velocity is given by $[(J_1+J_2)/4+2J_1J_2]^{1/2}$ 
in the former \cite{Matsuda} 
and $[-(J_1+J_2)/2]^{1/2}$ in the latter. 

\begin{center}
\psbox[scale=0.5]{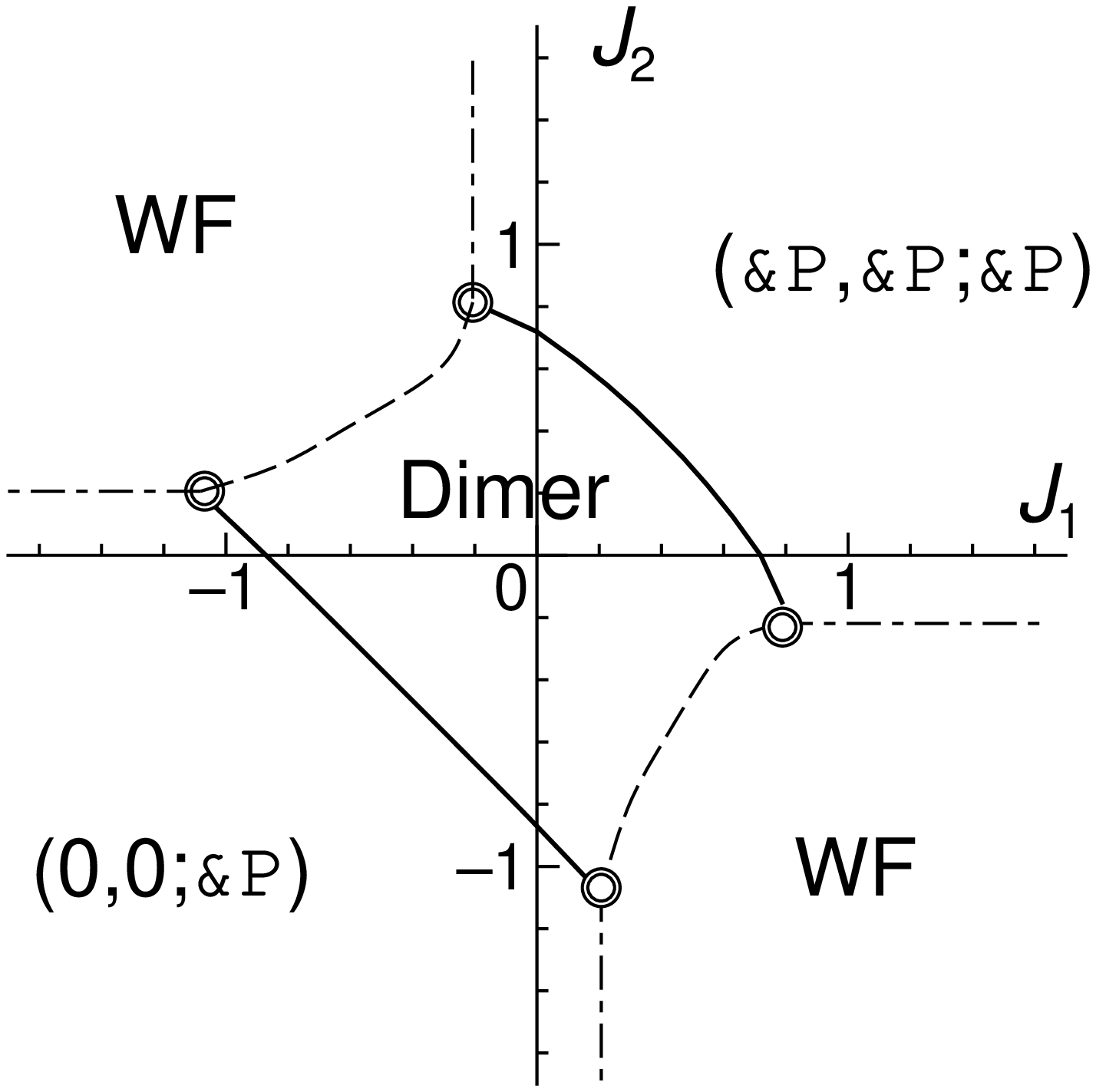}
\begin{quote}
\begin{footnotesize} 
Fig.~5. \hspace{6pt} Phase diagram 
for $S=1/2$ spins based primarily on series expansions. 
Solid lines represent precise estimates of continuous transitions; 
the dashed lines are more uncertain estimates 
of transitions, whose characters are discussed in the text. 
The dot-dashed lines are based on finite-size calculations. 
\end{footnotesize}
\end{quote}
\end{center}
\vspace{10pt}

In region III, the situation is more complicated
than in the preceding two. 
Based on the 
classical phase diagram and dimer mean-field theory,
one expects 
that the dimer-$(\pi,\pi;\pi)$ and 
dimer-$(0,0;\pi)$ critical lines 
would extend somewhat into this region, and this is
in fact observed.
On increasing $\lambda$, a dimer-$(\pi,\pi;\pi)$ transition takes place 
at $\theta=-\pi/16$, and likewise 
a dimer-$(0,0;\pi)$ transition occurs for $\theta=-7\pi/16$: 
the relevant gaps vanish and susceptibilities diverge
at the same value of $\lambda$ with the expected exponents. 

However, for most values of $\theta$ one would anticipate
encountering dimer-WF phase transitions. 
In fact there is an evidence for such transitions 
from unbiased approximants to the series, but the detailed features 
are rather complicated, in that neither the first-order 
nor continuous phase transition scenario receives 
complete support. 
Along the dashed line in Fig.~5, 
one finds that $\epsilon(\pi,\pi)$ vanishes 
with exponents close to 1/2.
The staggered susceptibility diverges with
exponents in the range 0.8--1.0.  So far, this is consistent
with the continuous transition scenario.  However, the 
critical values of $\lambda$ for the susceptibility are typically
20\% larger than those for the gap, and the 
``error bars'' (deduced from the consistency of 
the unbiased approximants) of two sets of critical 
points do not overlap with each other. 
Since the exponents of susceptibility 
are generally smaller than theoretically anticipated, one 
might hope that a biased analysis of the susceptibility would bring
the two sets of estimated critical points into agreement.
Unfortunately, this is not the case and 
biasing actually drives them farther apart. 
Consequently, we cannot rule out the possibility that the
dimer-WF transition involves a vanishing gap but finite susceptibility,
that is, a first-order transition.

\begin{center}
\psbox[scale=0.4]{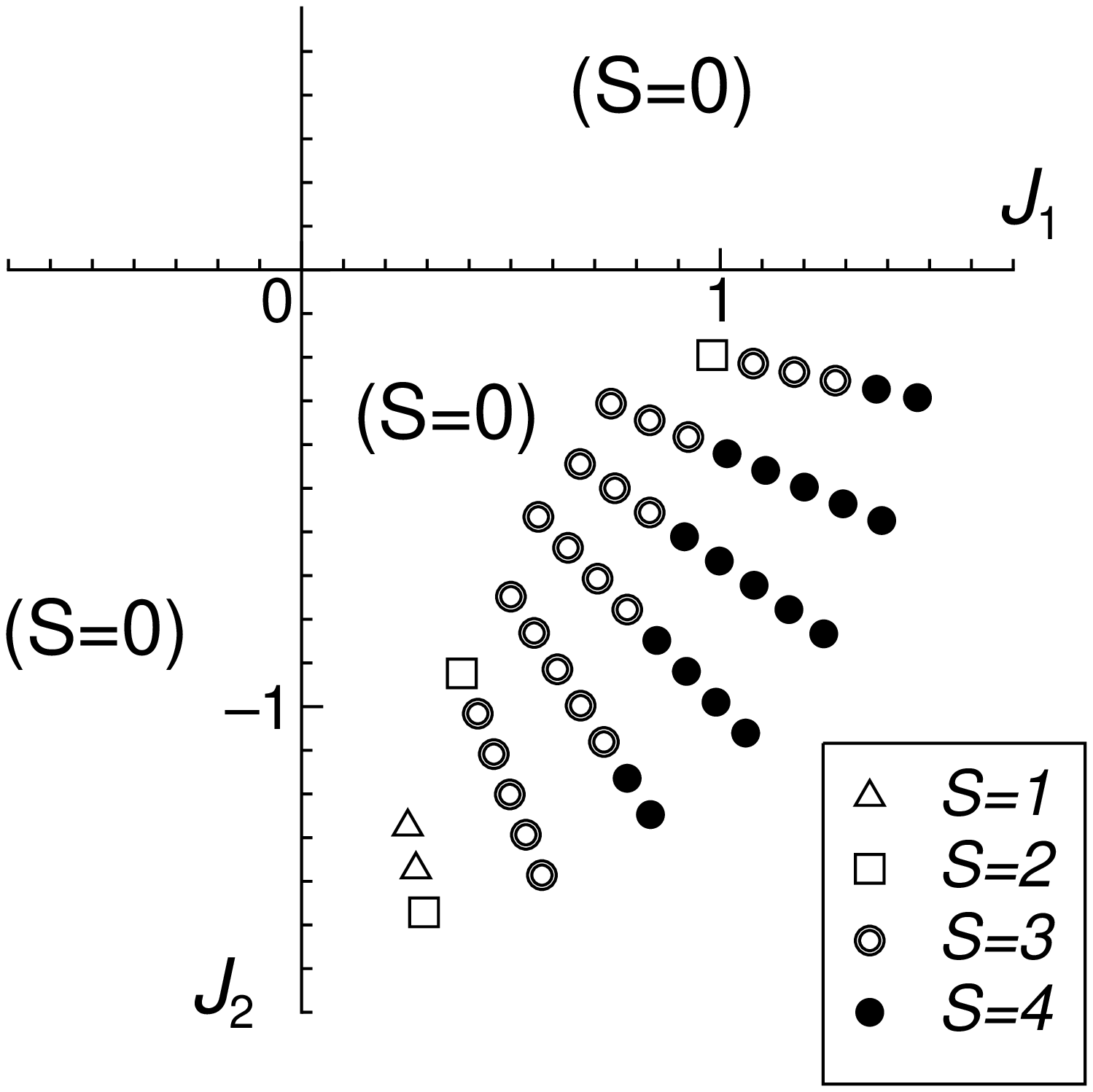}
\begin{quote}
\begin{footnotesize} 
Fig.~6. \hspace{6pt} A map of the total spin of 
ground state of $2\sqrt2{\times}2\sqrt2{\times}2$ 
clusters in the $J_1$-$J_2$ plane. 
\end{footnotesize}
\end{quote}
\end{center}
\vspace{10pt}
 
 As for the phase boundaries between 
the WF and magnetically ordered phases, the series 
expansions provide no information. 
The mean-field and classical calculations suggest that
they lie nearly parallel to the $J_1$ and $J_2$ axes.
In order to obtain a crude estimates of those phase
boundaries, we have constructed a map of the total spin of 
ground state of $2\sqrt2\times2\sqrt2\times2$ clusters 
in the $J_1$-$J_2$ plane, determined by exact diagonalization, 
see Fig.~6. 
The region where the ground state has nonzero spin is presumed
to lie in the WF phase.
The dimer-WF boundary estimated in this way
is consistent with that derived from the
series analysis. 
Therefore we expect that 
the boundaries between the WF and magnetically 
ordered phases (shown by the dot-dashed lines in Fig.~5) 
are reasonably accurate. 

At this point let us consider the properties of 
the excited states of the $2\sqrt2\times2\sqrt2\times2$ clusters 
in more detail. 
We find that in the vicinity 
of the singlet-singlet (dimer-to-antiferromagnetically ordered)
phase transitions the lowest excitation energy 
in the $S=2$ sector is twice greater than that in the 
$S=1$ sector, while in the vicinity of the
dimer-WF phase boundary the reverse is true: see Fig.~7. 
\begin{center}
\psbox[scale=0.35]{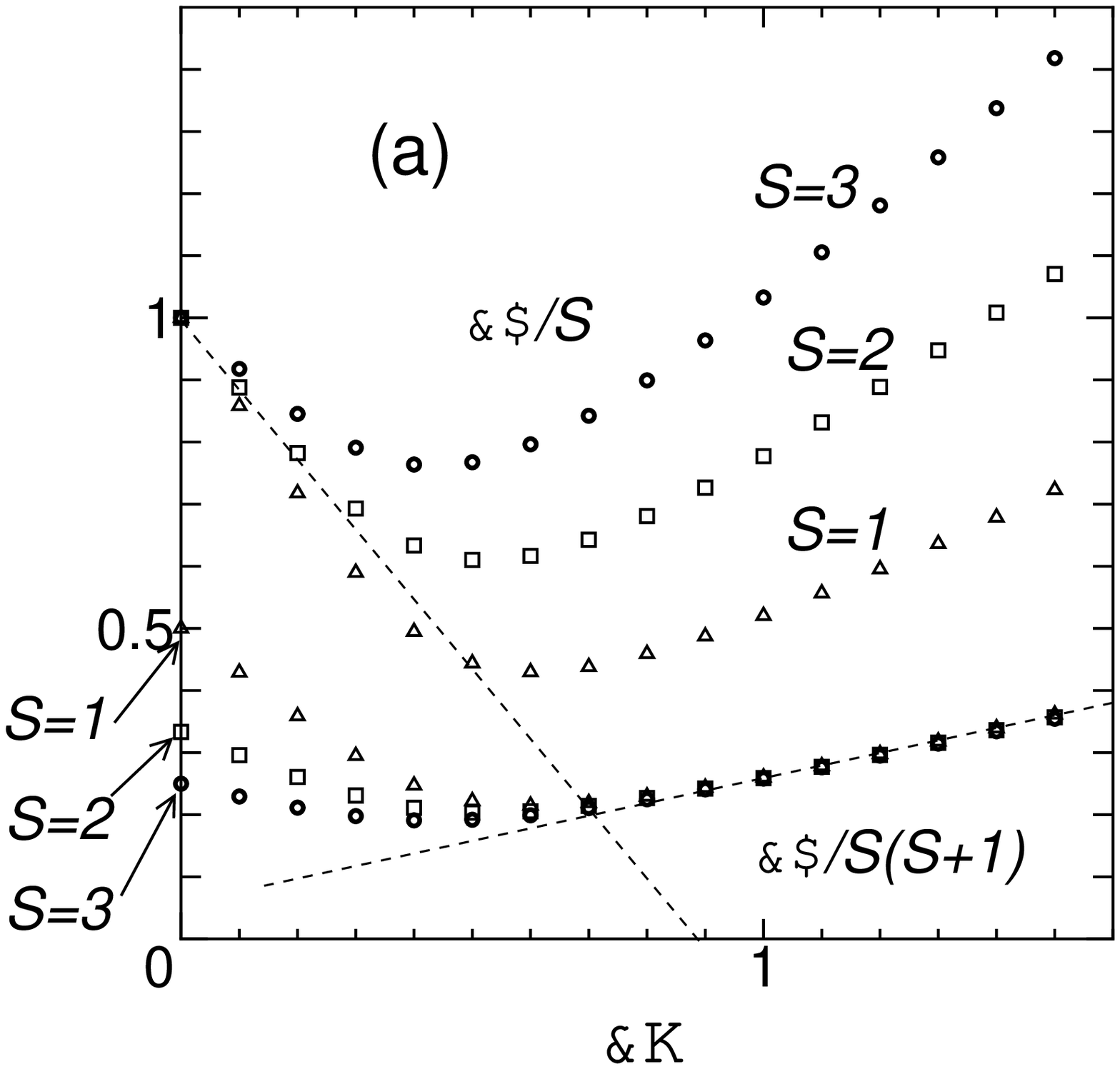}
\psbox[scale=0.35]{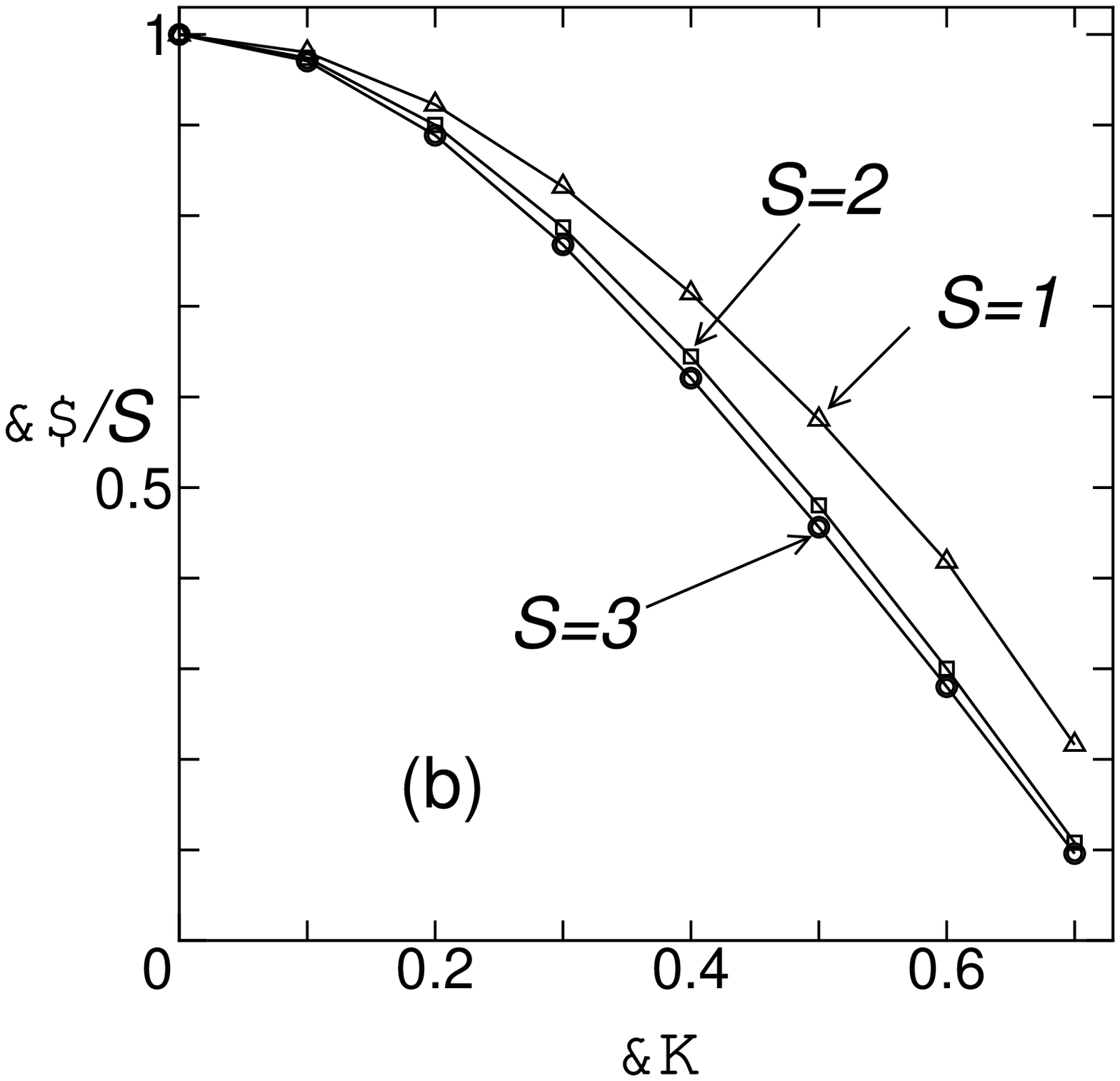}
\begin{quote}
\begin{footnotesize} 
Fig.~7. \hspace{6pt} Excitation gaps 
($\Delta =E_{\rm min}(S)-E_{\rm min}(S=0)$) 
scaled by $S$ or $S(S+1)$ 
in $2\sqrt2{\times}2\sqrt2{\times}2$ 
clusters as functions of $\lambda$:
(a) $\theta=\pi/4$, (b) $\theta=-\pi/4$. 
\end{footnotesize}
\end{quote}
\end{center}
\vspace{10pt}
Thus the triplet excitations are repulsive in the former cases
and attractive in the latter, as anticipated.
We have also examined how the lowest excitation energies in
each spin sector vary with $S$ in each of the singlet ground states. 
In the spin-disordered dimer phase those energies
are roughly proportional to $S$, while in the ordered phases
they are nearly proportional to $S(S+1)$.  Both results are
readily understood.  In the dimer phase the lowest spin-$S$ state
is essentially composed of $S$ triplet excitations. 
In the magnetically ordered phases 
the situation is rather different. 
In the bulk, arbitrarily low-energy excitations 
of any size of spins are possible, 
but for large but finite clusters 
the excitations amount to finite angular momentum 
states of the staggered moment (for which the corresponding 
moment of inertia is the uniform susceptibility) 
which accounts for the $S(S+1)$ scaling\cite{Bernu}. 
The crossover between the two different behaviors is impossible 
to pin down the critical points precisely 
in a system as small as the one 
we have considered, but the critical points obtained from the 
series analysis are clearly in the range where the crossover 
is taking place, see Fig.~7 (a). 

Finally let us present further results 
obtained from the series expansions 
regarding properties along the boundary of the dimer phase. 
The spin-wave velocity on the 
dimer-$(\pi,\pi;\pi)$ and dimer-$(0,0;\pi)$ critical lines
was determined
using the method described in Ref.~\citen{Gelfand1}. 
In Fig.~8, we show how the quantum renormalization factor 
of the critical spin-wave velocity 
defined by 
\begin{equation}
Z_{c}(\theta)=c/c_{\rm LSWT} 
\end{equation}
depends on $\theta$, where $c_{\rm LSWT}$ 
is the spin-wave velocity within linear spin-wave theory 
whose value was given earlier. 
To a good approximation we find that 
$Z_{c}(\theta)$ is independent of $\theta$ along 
the dimer-$(0,0;\pi)$ boundary. 
This seems to be consistent with the fact 
that the phase boundary itself can be well described 
by $J_1+J_2={\rm const}$. 
\begin{center}
\psbox[scale=0.4]{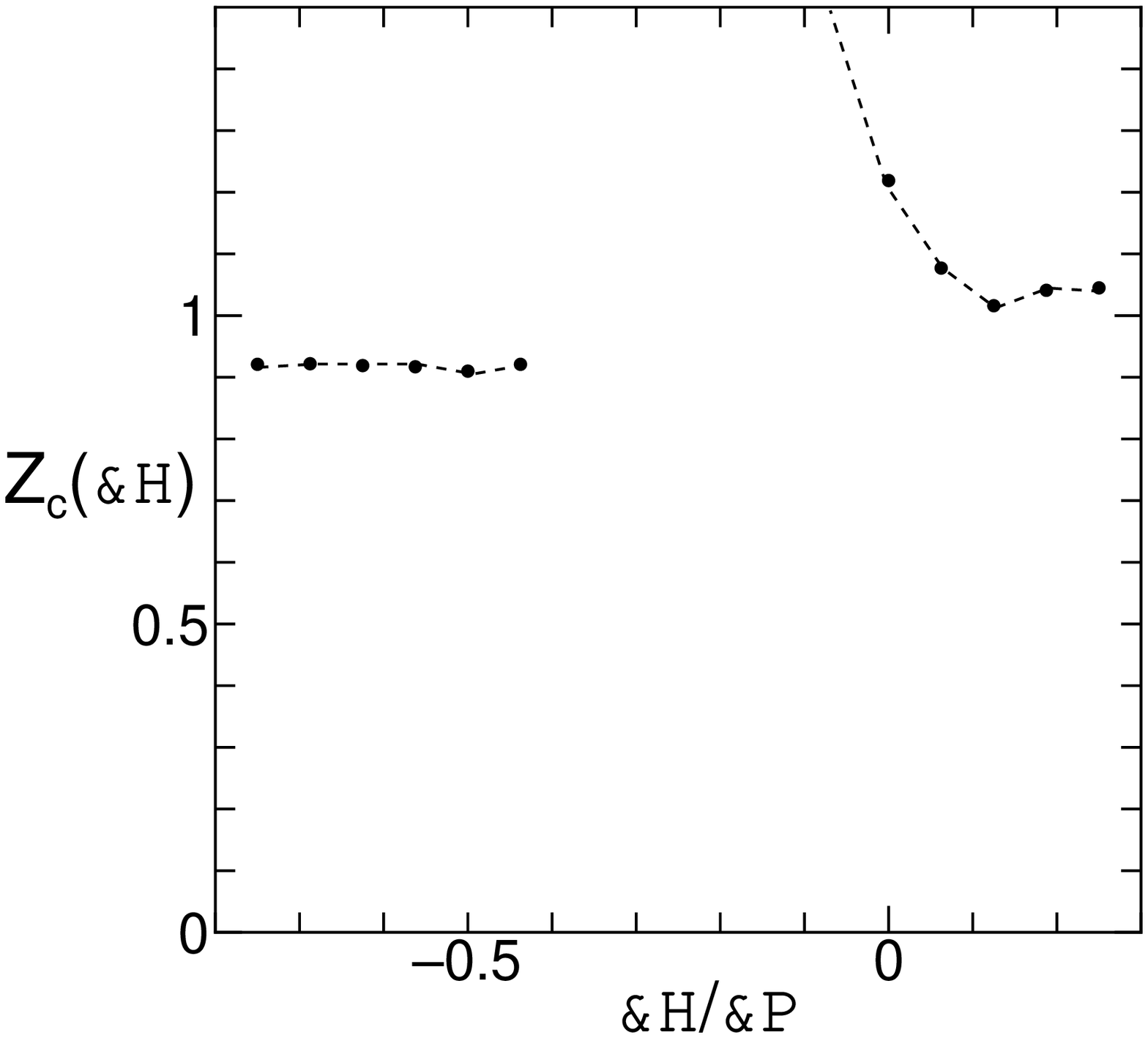}
\begin{quote}
\begin{footnotesize} 
Fig.~8. \hspace{6pt} Quantum renormalization factor 
of the critical spin-wave velocity $Z_{s}(\theta)$ along 
continuous transition boundaries as a function of $\theta$. 
\end{footnotesize}
\end{quote}
\end{center}
\vspace{8pt}

 One can also extrapolate the 
excitation spectra 
for values of ${\bf q}$ far from the 
characteristic momenta $(0,0)$ or $(\pi,\pi)$, 
where one expects direct Pad\'e approximants to 
provide accurate estimates. 
Estimates of the dispersion relations 
along selected lines in the Brillouin zone 
are presented for several $\theta$ values in Fig.~9. 

\begin{center}
\psbox[scale=0.4]{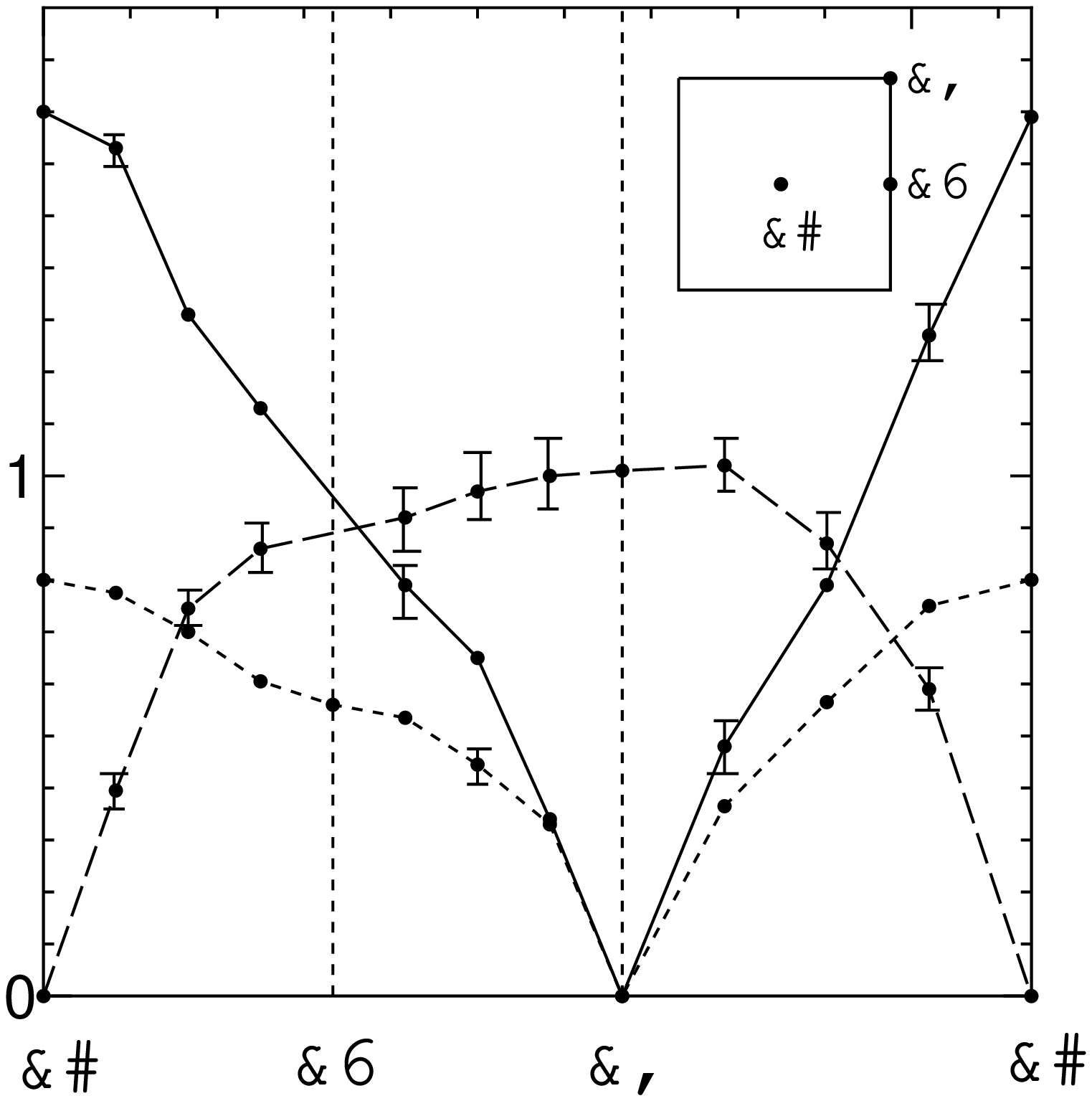}
\begin{quote}
\begin{footnotesize} 
Fig.~9. \hspace{6pt} 
Critical spin-wave excitation spectra: 
solid, dotted and dashed lines are 
for $\theta=0$, $\theta=-5\pi/16$ and 
$\theta=-\pi/2$, respectively. 
\end{footnotesize}
\end{quote}
\end{center}
\vspace{10pt}

\section{Summary and Discussion}

We have investigated the ground state properties 
and phase diagram of bilayer Heisenberg model 
on the square-lattice with 
general nearest-neighbor intralayer couplings. 
The ground state phase diagram of the model has been 
first examined in the classical limit 
in order to get an insight to its 
asymptotic behaviors. 
Then the effects of quantum fluctuations are studied 
on the basis of several methods; 
dimer mean-field theory, small cluster 
diagonalization, and high-order perturbation theory 
about the interlayer dimer limit. 
We have found four phases for the model with $S=1/2$: 
one ``spin liquid'' (dimer) phase which is the analytic continuation 
of the interlayer dimer Hamiltonian, two antiferromagnetically
ordered [$(\pi,\pi;\pi)$ and $(0,0;\pi)$] 
phases, and one phase with weak magnetization stabilized by 
competing interactions between layers (WF). 
The transitions between the dimer phase and the antiferromagnetically
ordered phases are continuous 
(and belong to the classical $d=3$ Heisenberg 
model universality class), while the character of the
dimer-WF transitions could not be reliably determined.
Concerning the dimer-$(\pi,\pi;\pi)$ and dimer-$(0,0;\pi)$ 
transitions, it appears that quantum fluctuations
play a more important role in the former, and lead to a rather 
intriguing result that
N\'eel order is suppressed more strongly
when the intralayer couplings are
evenly distributed between the layers than when they
are concentrated in one layer. 

The case ($\theta=0$) is especially interesting. 
In this case, the spins are antiferromagnetically coupled 
in one of layers and not coupled at all in the other. 
This model corresponds to symmetric Hubbard-Kondo square-lattice model 
in the large-$U$ limit. Provided $t\to\infty$ and 
$U\to\infty$ with $4t^2/U=J_s={\rm constant}$ 
(the intralayer coupling) 
and the on-site Kondo coupling (the interlayer 
coupling) is $J_K$, we have found that 
there is a transition to a magnetically ordered state 
when $J_s$ is increased up to approximately $0.71 J_K$. 
This leads to the result for the 
critical $t$ mentioned in the Introduction.

The one-dimensional
version of the Hamiltonian (1.1), with general intrachain couplings,
has been attracting interest recently in connection with 
the spin-ladder materials\cite{Dagotto}.  
In particular, Tsukano and 
Takahashi \cite{TT} studied the one-dimensional model 
with $J_2=-J_1$ ($\theta=-\pi/4$) 
and found a continuous transition 
between the dimer phase and a magnetically ordered phase, 
supporting continuous transition scenario of the 
dimer-WF transition. 

Finally, Sachdev and Senthil \cite{Sachdev} have studied a rather
general quantum rotor model for which, in the
large-$S$ limit, there is a direct mapping to the model (1.1).
It is interesting to see which features of their model
persist for $S=1/2$, and which do not.  The mean-field analysis
of the rotor model finds a ``gapped quantum paramagnet,''
equivalent to the dimer phase in the spin model, a N\'eel
phase, a ``quantized ferromagnet'' phase, and 
a canted phase, equivalent to the WF phase in the spin model.  
They suggest that, quite generally, continuous phase transitions into 
the canted phase, where there is nonzero magnetization, are only possible from
phases with gapless excitations.   Consequently, continuous phase
transitions between the dimer phase and the WF phase should not
be possible.  (In the section of the phase diagram for the rotor
model which is presented by Sachdev and Senthil, not even first
order transitions between the gapped quantum paramagnet and canted
phases are exhibited, but there is no reason to believe that holds
for all choices of parameters.\cite{SSpc})
Our numerical studies were not able to establish the nature of
the dimer-WF transition, and so our results are not in direct
disagreement with the general principle espoused by Sachdev
and Senthil even though they do not provide any support for it, either.  
However, there is one significant difference
between the phase diagrams of the rotor model and of the spin model.
For $S=1/2$, the Hamiltonian (1.1) does not exhibit an
analog of the quantized ferromagnet phase.  {\it Ferromagnetically\/}
coupled bilayers of $S=1/2$ spins could exhibit such a phase,
since in the absence of intralayer coupling the interlayer dimers
would have angular momentum 1.
In contrast to the rotor model, however, it does not appear possible
to go to such a phase directly from the dimer phase by varying the
intralayer couplings.

\vspace{1cm}

\section*{Acknowledgements}
We thank S. Sondhi for discussions.
This work has been supported by the U. S. National Science
Foundation through grant DMR 94--57928 (MPG).


\begin{thebibliography}{99}
\bibitem{Nishimori} The numerical calculations were 
done with the use of the program TITPACK Version 2 
developed by H. Nishimori at the Tokyo Institute of  Technology.
\end{thebibliography}
\end{document}